# Second order anisotropy contribution in perpendicular magnetic tunnel junctions

A.A. Timopheev[1,2,3*], R. Sousa[1,2,3], M. Chshiev[1,2,3], T. Nguyen[1,2,3] and B. Dieny[1,2,3]

1. Univ. Grenoble Alpes, INAC-SPINTEC, F-38000 Grenoble, France
2. CEA, INAC-SPINTEC, F-38000 Grenoble, France
3. CNRS, SPINTEC, F-38000 Grenoble, France

e-mail: *andrey.timopheev@gmail.com*

Magnetoresistance loops under in-plane applied field were measured on perpendicularly magnetized magnetic tunnel junction (pMTJ) pillars with nominal diameters ranging from 50 to 150 nm. By fitting the hard-axis magnetoresistance loops to an analytical model, the effective anisotropy fields in both free and reference layers were derived and their variations in temperature range between 340K and 5K were determined. It is found that an accurate fitting is possible only if a second-order anisotropy term of the form $-K_2 \cos^4\theta$, is added to the fitting model. This higher order contribution exists both in the free and reference layers and its sign is opposite to that of the first order anisotropy constant, $K_1$. At room temperatures the estimated $-K_2/K_1$ ratios are 0.1 and 0.24 for the free and reference layers, respectively. The ratio is more than doubled at low temperatures altering the ground state of the reference layer from "easy-axis" to "easy-cone" regime. Easy-cone state has clear signatures in the shape of the hard-axis magnetoresistance loops. The same behavior was observed in all measured devices regardless of their diameter. The existence of this higher order anisotropy was confirmed experimentally on FeCoB/MgO sheet films by ferromagnetic resonance technique. It is of interfacial nature and is believed to be linked to spatial fluctuations at the nanoscale of the anisotropy parameter at the FeCoB/MgO interface, in agreement with Dieny-Vedyayev model.

1. Introduction

Magnetic anisotropy is a key feature of a ferromagnetic material playing a crucial role in technical applications of these materials. Generally, this phenomenon takes its origin from magnetic dipole-dipole, exchange and/or spin-orbit interactions. These interactions provide respectively shape, exchange and magnetocrystalline (magnetoelastic) anisotropies. One can also divide the magnetic anisotropy as arising from the bulk and/or from the surface or interface of the layer.

Concept of interfacial anisotropy was proposed in the pioneering work of L. Neel [1] predicting the perpendicular interfacial anisotropy as a result of the lowered symmetry at the surface/interface. This work was followed by experiments carried out on ultrathin NiFe films grown on Cu(111) [2] which confirmed the interfacial nature of the perpendicular magnetic anisotropy (PMA) observed in this system. Within the last fifty years a lot of work has been carried out on interfacial anisotropy both from theoretical and experimental points of view [3-8]. Nowadays, perpendicular interfacial anisotropy has become one of the main ingredients of novel magnetic memory elements employing out-of-plane magnetized (perpendicular) magnetic tunnel junctions (pMTJ) stacks [9-11]. In such structures, perpendicular anisotropy of the free layer is provided by the interface between FeCoB and MgO layers while in the reference layer, it is additionally enhanced by exchange coupling with Co/Pt or Co/Pd multilayers [12] with PMA of interfacial nature as well.

Taking into account the system symmetry, the PMA energy density originating from the interface can be written as:

$$E_{PMA} = -\frac{1}{t}(K_{1s}\cos^2\theta + K_{2s}\cos^4\theta + \cdots), \qquad (1)$$

where θ is the angle between magnetization and normal to the plane of the layers, $K_{1s}, K_{2s}$ … are constants of the first and second order surface anisotropy energy per unit area and $t$ is the thickness of the FM layer. One can then define effective bulk anisotropy constants which also include the demagnetizing energy for a thin film (CGS units) : $K_1 = (\frac{K_{s1}}{t} - 2\pi M_S^2)$, $K_2 = \frac{K_{s2}}{t}$. In case of very thin Fe films magnetization saturation parameter $M_S$ is typically reduced in comparison with its bulk value [13]. If $K_1 > 0, K_2 < 0$ and $0.5 < {-K_2}/{K_1} < 1$, the ground state of the system will correspond to so-called "easy-cone" regime, or canted state. In the easy-cone regime, the magnetization is tilted away from the symmetry axis by angle $\theta_c$ defined according to $\cos^2\theta_c = {-K_1}/{2K_2}$ . Due to the axial symmetry, the system energy remains invariant around a cone with opening angle $\theta_c$ yielding a so-called "easy-cone" anisotropy. Quite frequently, in systems where interfacial anisotropy is present, the first order term proportional to $K_{1s}$ dominates the higher order term proportional to $K_{2s}$ . However, the influence of this second order term has been clearly observed experimentally around the spin-reorientation transition region where the demagnetizing energy partially or fully balances the $K_{1s}\cos^2\theta$ term (i.e. effective anisotropy $K_1$ is close to zero) [14-19]. $K_{2s}\cos^4\theta$ term can arise due to peculiarities of atomic structure at the interface or as a result of non-uniform mechanical stresses existing at interfaces presenting a large crystallographic mismatch. Also, B. Dieny and A. Vedyayev have shown analytically that spatial fluctuations of the film thickness under $K_{1s} = const$ term can lead to a higher order $K_{2s}\cos^4\theta$ term if the period of the fluctuations is lower than the exchange length of FM material [20]. Recently, J. Sun has reported similar results [21].

Experimental determination and understanding of magnetic anisotropy in FM layers and multilayers is very important towards the pMTJ stack optimization for future use in STT-MRAM applications. Experiments conducted on sheet films combining magnetometry (VSM, SQUID etc.) with ferromagnetic resonance (FMR) allow the determination of magnetic anisotropy constants at sheet film level. However, in the context of STT-MRAM development, it is also important to know how these anisotropy parameters are affected by the patterning process and how they are distributed from dot to dot in an array of magnetic tunnel junctions. Magnetotransport measurements with field applied in the plane of the layers provide a convenient way to determine the anisotropy characteristics in pMTJ. Magnetic field applied along hard-axis tilts the magnetic moments of both layers away from the normal to the plane direction which produces a change in the tunneling conductance of the system. The curvature of the obtained MR(H) dependences and their different shapes for initially parallel and antiparallel magnetic configurations allow direct extraction of the effective anisotropy fields in both magnetic electrodes assuming that micromagnetic distortions are not developing much under the applied field (macrospin approximation). Such analysis can be performed on an automated wafer prober equipped with an electromagnet allowing large-scale analysis of pMTJ pillar arrays with good statistics. For deeper analysis on a limited number of pMTJs, experiments can also be carried out on experimental setups such as Physical Property Measuring System (PPMS) allowing measurements in a wide range of temperatures and magnetic fields.

In this study, we investigated the anisotropy in pMTJs via hard-axis magnetoresistance loops analysis and derived the effective anisotropy fields of these pMTJ pillars of various nominal diameters ranging from 50nm to 150 nm. The 1st and 2nd order magnetic anisotropy in both layers were derived as well as their temperature dependences. It was found that a significant $K_{2s}\cos^4\theta$ term is present in both free and polarizing layers. This term has a negative sign of $K_{2s}$ and can result in an easy-cone magnetic state with canted remanence of the magnetic layers.

## 2. Experimental details

pMTJ pillars array with nominal diameters ranging between 50nm and 500 nm were fabricated from an MTJ stack grown by magnetron sputtering. The stack contains a 1.7nm thick $Fe_{60}Co_{20}B_{20}$ free layer sandwiched between two MgO barriers. Saturation magnetization parameter of the free layer was measured to be 1030 emu/cm$^3$. Current in-plane magnetotransport measurements yielded RxA = 5.7 Ω μm$^2$ and TMR=126 %. The second MgO barrier was introduced to increase the perpendicular anisotropy of the free layer. It has a negligible resistance-area (RA) product compared to the main tunnel barrier. The bottom reference layer is synthetic antiferromagnetic (SAF) and comprises perpendicularly magnetized (Co/Pt) multilayers and a polarizing material next to the MgO barrier which has the same composition as that of the free layer. The metallic electrode above the second MgO barrier is non-magnetic. Additional information and experiments on these samples can be found in ref. [22].

Statistical measurements of coercivity, coupling field and TMR values were performed using an automated wafer prober setup equipped with an electromagnet. Temperature-dependent measurements on single pMTJ pillars were carried out using PPMS system. Magnetoresistance loops were measured by applying a magnetic field along the easy and hard axis directions and passing a constant current through the pillars which amplitude was set not to exceed 30 mV voltage drop across the tunneling barrier in the antiparallel configuration in order to minimize any spin-transfer-torque influence during the measurement. At each field point, the voltage drop was measured and the resistance determined. Magnetic field was swept from -6 kOe to +6 kOe and then back to -6 kOe with a constant sweep rate.

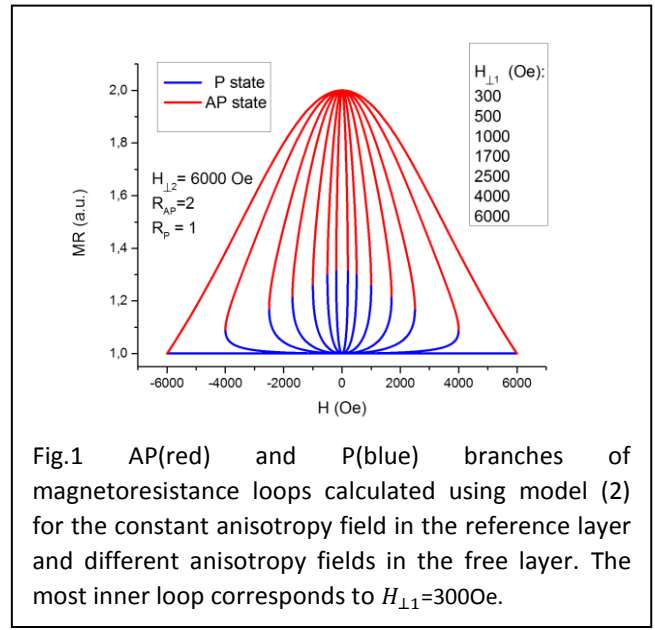

Fig.1 AP(red) and P(blue) branches of magnetoresistance loops calculated using model (2) for the constant anisotropy field in the reference layer and different anisotropy fields in the free layer. The most inner loop corresponds to $H_{\perp 1}$=300Oe.

## 3. Analytics of hard-axis magnetoresistance loops

Assuming $K_1 \neq 0$, $K_2 = 0$, macrospin behavior and linear dependence of the tunneling conductance versus cosine of the relative angle between magnetization vectors in the two magnetic electrodes [23], one can analytically derive the hard-axis magnetoresistance as a function of applied field H for initially (at H=0) parallel and antiparallel states:

$$MR(H) = 2 R_P R_{AP} \left( R_P + R_{AP} + (R_{AP} - R_P) \left( \pm \sqrt{\left(1 - H^2/H_{\perp 1}^2\right)\left(1 - H^2/H_{\perp 2}^2\right)} + H^2/(H_{\perp 1} H_{\perp 2}) \right) \right)^{-1}, \quad (2)$$

where $H_{\perp 1}$, $H_{\perp 2}$ are the effective perpendicular anisotropy fields in the two electrodes, $R_P, R_{AP}$ are the resistance values in parallel and antiparallel state, plus/minus sign of the square root corresponds to MR curve for the initially parallel/antiparallel state.

Fig.1 shows the variation of MR curves defined by Eq. (2) starting from the P or AP states with respect to $H_{\perp 1}/H_{\perp 2}$ ratio. When $H_{\perp 1}/H_{\perp 2} \ll 1$, both curves starting from P or AP states have parabolic shape with similar curvatures. This behavior corresponds to the limit of strictly fixed reference layer. On the contrary, if both layers have the same anisotropy fields, $H_{\perp 1} = H_{\perp 2}$, the resistance starting from P state will remain unchanged whatever the field (both magnetization rotates together), while the MR curve starting from the AP state will vary from $R_{AP}$ to $R_P$ value. The variation of the curvatures with respect to $H_{\perp 1}/H_{\perp 2}$ ratio allows one to estimate $H_{\perp 1}$ and $H_{\perp 2}$ directly from the experiments by fitting the experimental hard-axis MR curve starting from P and AP states with expression (2). Knowing the magnetization saturation parameter and ferromagnetic film thickness, one can derive the surface anisotropy constant $K_s$ from the relation: $\frac{H_\perp M_S}{2} = (\frac{K_s}{t} - 2\pi M_S^2)$. If higher order anisotropy contributions have to be taken into account in (1) in order to improve the fits, then no analytical expression similar to Eq. (2) is available but the fitting of the MR(H) curves is still numerically feasible.

One should notice, however, that micromagnetic distortions, strong interlayer coupling and superparamagnetic thermal fluctuations can play a role in the MR(H) dependences and worsen significantly the fitting quality. Analysis of easy-axis MR(H) loops can help to identify possible contributions of these effects.

### 4. Room-temperature easy-axis magnetoresistance loops

Using the automated wafer prober setup, about 90 pillars of each diameter were measured to obtain statistically reliable information and to guide the choice of the samples for a further more detailed investigation of the anisotropy properties. 15-loop magnetoresistance hysteresis loops were

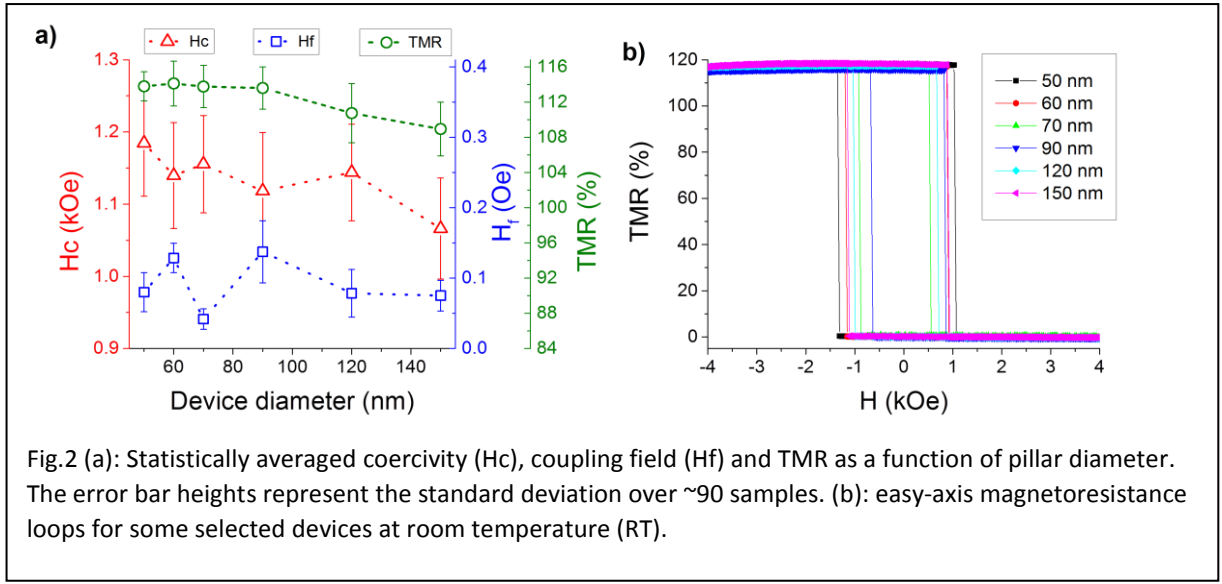

Fig.2 (a): Statistically averaged coercivity (Hc), coupling field (Hf) and TMR as a function of pillar diameter. The error bar heights represent the standard deviation over ~90 samples. (b): easy-axis magnetoresistance loops for some selected devices at room temperature (RT).

measured on each device. The magnitudes of the TMR, coercive field and coupling field were extracted from the averaged loops. Few devices showing TMR < 90% were excluded from the statistical analysis. Fig. 2(a) shows these three parameters as a function of pillar diameter. In average, all samples have a coercive field ~1.1 kOe, a coupling field ~ 90 Oe and TMR ~113%. For most devices, the interlayer exchange coupling is ferromagnetic with a positive sign. It is hard to track its diameter dependence since the standard deviation is of the same order of magnitude as the mean measured value. It is believed that fluctuations of the coupling field are mainly caused by damages of the pillar edges. No correlations are observed between $H_f$, $H_c$ and TMR values. The coercivity and TMR are observed to weakly decrease versus pillar diameter, which can be ascribed to the appearance of micromagnetic distortions at pillar edges as the diameter increases (e.g. flower state). Individual easy-axis magnetoresistance loops of some selected devices at RT are shown in Fig. 2(b). The measurements were performed on a PPMS-based setup at room temperature. All measured devices have similar TMR amplitude and perfect rectangular shape with no evidence of any intermediate states between full P and AP configurations.

### 5. Temperature dependent measurements

In a hard-axis measurement of the MR(H) loops, the magnetization of the storage layer only rotates by 90° between the remanent state and the saturated state. The system has to be prepared either in the initial P configuration or in the initial AP configuration giving two hard-axis hysteresis branches. If the field is strictly applied in the plane of the sample during the hard-axis measurement, the two MR(H) branches can be obtained according to the following protocol with 8 steps using a PPMS setup with rotating sample holder : 1) switch the pMTJ pillar in P state by applying the magnetic field along the easy axis, set H=0 and rotate the pillar into hard-axis configuration; 2) make a MR(H) measurement from H=0 to H=Hmax; 3) repeat step 1.; 4) make a MR(H) measurement from H=0 to H= -Hmax; 5) rotate the sample back to the position with field applied parallel to the normal to the plane (i.e. along easy axis) and set the pillar in AP state, set H=0 and rotate the pillar into hard-axis configuration; 6) repeat step 2; 7) repeat step 5; 8) repeat step 4. By putting together MR(H) dependences obtained for negative and positive magnetic field sweeps, one finally obtains the two MR(H) branches corresponding to initially P and AP states, i.e. a full hard-axis MR loop.

We have implemented a simplified method for hard-axis MR(H) measurements. If the magnetic field is slightly tilted away from the hard axis by a few degrees, then the small out-of-plane remaining component of the applied field will allow the switching of the magnetization from the P hard-axis branch to the AP hard axis branch. Thermal fluctuations and interlayer exchange coupling across the tunnel barrier determine the minimal angle of misalignment necessary to observe these jumps between the two branches. In our samples where the coupling field is one order of magnitude lower than the switching field, it is enough to tilt the magnetic field by 3-4 degrees out-of-plane which slightly distorts the MR(H) curves, making them slightly asymmetric around the vertical axis. But at the same it allows obtaining the full hysteresis loop containing both AP and P branches much more easier than the case when the field is applied strictly in-plane. Here and further we will call AP/P branches those corresponding to the reversible parts of a MR(H) hysteresis loop with respective AP/P state at H=0. As an example, let us describe a MR(H) loop measured on 70 nm pMTJ pillar at T=340K (the most inner loop in Fig.3). The MR(H) loop contains both AP and P branches both having a parabolic shape before the switching occurs. The AP (resp. P) branch has a maximum (resp. minimum) at H=0 with R=5.8 kΩ (resp. 2.81 kΩ). The switching fields between the branches (which are seen as vertical lines) are -1.7 kOe and +1.5 kOe for P->AP and AP->P branch transitions, respectively. The resistance range corresponding to a discontinuous change in magnetoresistance (the switching) is cut out from the graph in order to focus the reader attention on the reversible parts of MR(H) dependence situated in-between the switching fields and which is only discussed in the following of the text. Thus, the graph has a brake hiding a range between 3 and 5 kΩ and it has a different vertical scale before and after the brake due to noticeable difference in MR(H) curvature for P and AP branches. The same is applied below in Fig.4.

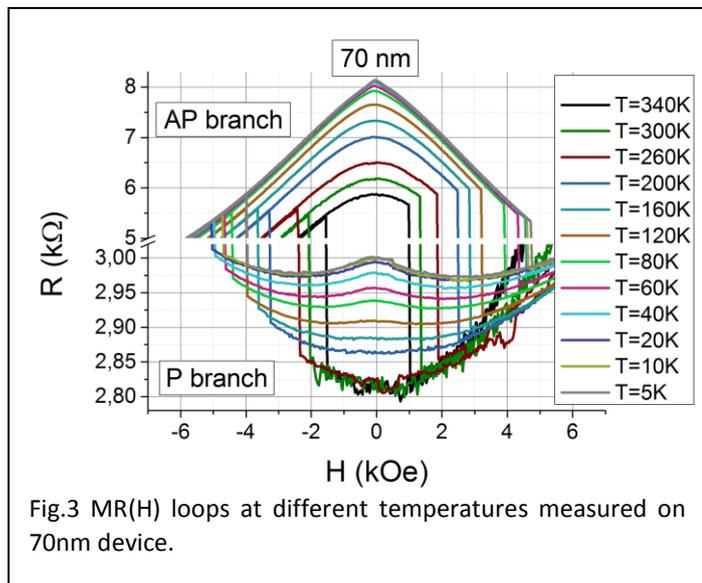

Fig.3 MR(H) loops at different temperatures measured on 70nm device.

Figure 3 shows MR(H) loops behavior as a function of temperature ranging between 5K and 340K for a 70nm diameter pMTJ pillar. For T > 140-120 K, it qualitatively reproduces the situation described in Section 3, i.e. both AP and P branches have a characteristic parabolic shape. In the AP state the curvature is more pronounced; the resistance variation for the AP branch is one order of magnitude larger than for P branch, which can be ascribed to the finite PMA of the reference layer and correlatively to a rotation of its magnetization. The fitting according to Eq. (2), however, is not ideal even at high temperatures and it is getting worse at decreasing temperature. For T<120K MR(H) loops gradually gain new qualitatively different features and it becomes impossible to reproduce the shape of AP and P branches using Eq. (2). Indeed, at T=5K AP branch exhibits a triangular shape while the P branch shows a local maximum of resistance at H=0 and two respective minima located at +/- 2.5-2.7 kOe. The same behavior is observed for all device diameters, as shown in Fig.4.

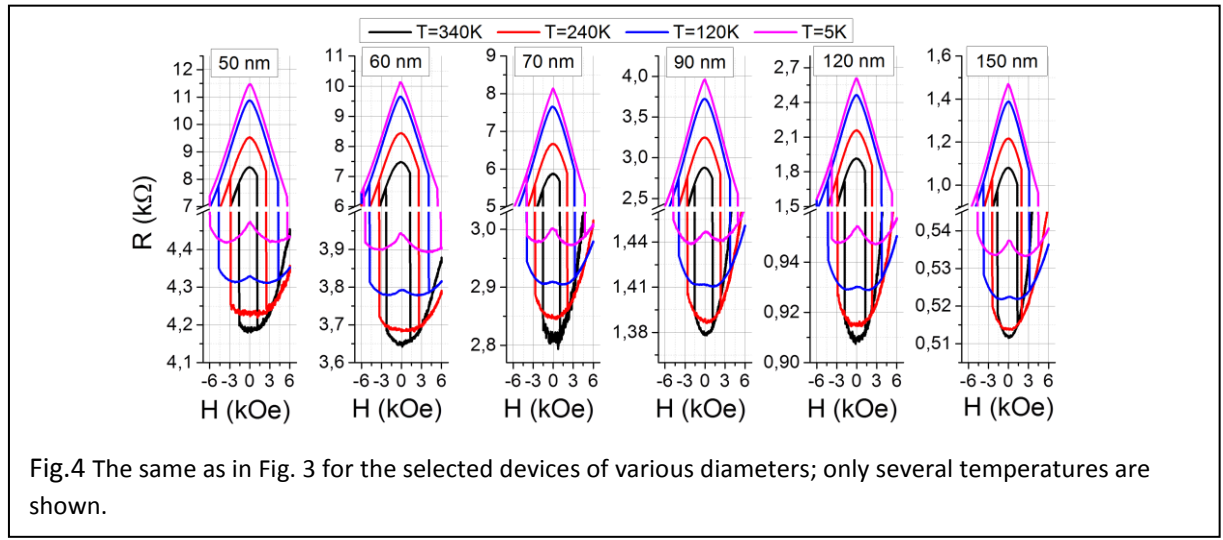

Fig.4 The same as in Fig. 3 for the selected devices of various diameters; only several temperatures are shown.

To reproduce experimentally the obtained results in a wide range of temperatures, the model giving Eq. (2) needs to be improved by introducing a second-order uniaxial anisotropy term both in the free and reference layers. The total magnetic energy density (normalized by magnetization saturation parameter $M_S$) in each layer can be then written as follows:

$$\frac{E_{tot}}{M_S} = -\frac{K_1}{M_S}\cos^2\theta - \frac{K_2}{M_S}\cos^4\theta - H\sin\theta, \qquad (3)$$

Each layer assumed to behave as a macrospin. Considering that the uniaxial anisotropy has an interfacial origin, the effective anisotropy constants can be written as $K_1 = (\frac{K_{s1}}{t} - 2\pi M_S^2)$, $K_2 = \frac{K_{s2}}{t}$, where $K_{s1}, K_{s2}$ are the interfacial perpendicular anisotropy constants, $t$ is the thickness of a layer.

Unfortunately, no analytical expression of the R(H) variation can be derived in this case. However, the fitting can be carried out numerically. In this case we folded up the AP and P branches around the H = 0 horizontal axis in order to have a more accurate fitting and to cancel, at least partially, the asymmetry of the left and right wings of the MR(H) dependences appearing due to tilted orientation of the external magnetic field. We also increased the relative magnitude of the P branch to give equal weight in the fitting procedure of the P and AP branches. Both AP and P branches are actually fitted simultaneously, so that each fitting result gives $\frac{K_1}{M_S}, \frac{K_2}{M_S}$ values in both free and reference layers. Typical results of the fit are shown in Fig. 5. The higher magnitudes set of $K_1, K_2$ corresponds obviously to the reference

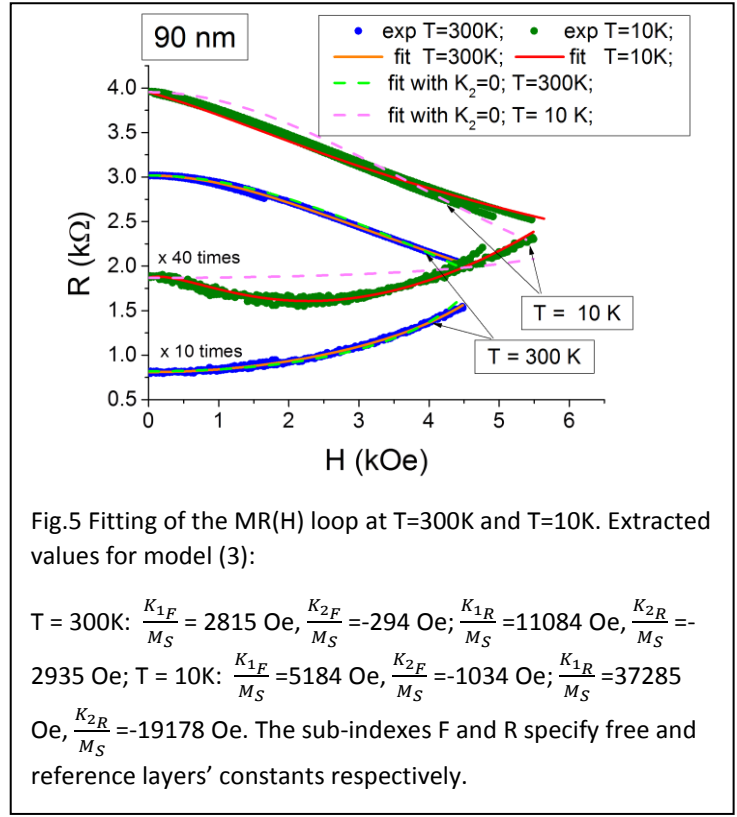

Fig.5 Fitting of the MR(H) loop at T=300K and T=10K. Extracted values for model (3):

T = 300K: $\frac{K_{1F}}{M_S}$ = 2815 Oe, $\frac{K_{2F}}{M_S}$ =-294 Oe; $\frac{K_{1R}}{M_S}$ =11084 Oe, $\frac{K_{2R}}{M_S}$ =-2935 Oe; T = 10K: $\frac{K_{1F}}{M_S}$ =5184 Oe, $\frac{K_{2F}}{M_S}$ =-1034 Oe; $\frac{K_{1R}}{M_S}$ =37285 Oe, $\frac{K_{2R}}{M_S}$ =-19178 Oe. The sub-indexes F and R specify free and reference layers' constants respectively.

layer. We will use "F" and "R" sub-indexes in the constants to specify to which layer these constants are associated. Accuracy of the fitting is very high in the temperature range between 340 and160K. At lower temperatures, the fitting is less accurate but still good enough to reproduce both the triangular shape of the AP branch and characteristic double-well shape of the P branch. It is also important to illustrate how the fitting with $K_{2F}, K_{2R} = 0$ (model Eq. (2)) looks like for the same experimental data (see Fig. 5, dotted lines). At high T=300K one can conclude that fitting according to Eq. (2) becomes acceptable. However even in this case, a deviation from the experimental curves is clearly observed: the obtained R(H) curvature is not as accurately reproduced as in the case where the fitting includes the second order anisotropy term. At low temperatures, the fitting without including the second order anisotropy terms does not work at all because of the impossibility to reproduce the double-well shape of the P branch.

The origin of the appearance of a triangular shape in the AP R(H) branch as well as double well shape in the P R(H) branch at low-temperature can be understood from the values of the extracted $K_1$ and $K_2$ parameters for T=10K. In both layers $K_{2F}, K_{2R}$ is negative. In the free layer $-K_{2F}/K_{1F}$ = 0.2 versus $-K_{2F}/K_{1F}$ = 0.104 at T=300K. Increase of $-K_{2F}/K_{1F}$ ratio at low temperatures results mainly in decrease of the free layer's switching field and deformation of hard-axis M(H) and R(H) dependences. In the case of the reference layer at T=300K $-K_{2R}/K_{1R}$ =0.265 while at T=10K $-K_{2R}/K_{1R}$ =0.514, i.e. higher than 0.5 which yields the onset of an "easy-cone" ground state of the reference layer magnetization instead of "easy-axis" at high temperatures. In the easy-cone regime, the reference layer magnetization is tilted out from the symmetry axis by angle $\theta_c$, $\cos^2 \theta_c = -K_1/2K_2$ ; From the fitting parameters, the easy-cone angle at 10K is $\theta_{c_R} \sim 10°$. In this regime, an infinitely small reversal of the in-plane applied field yields a 180° rotation of in-plane component of the reference layer magnetization around its easy cone thereby skipping the parabolic part of the R(H) curve, thus resulting in the observed triangular shape of the R(H) response at low temperature.

The double-well shape of the P branch can also be explained by the easy-cone regime in the reference layer. The free layer is in the "easy-axis" state (i.e. at H=0, $\theta = 0$) since $-K_{2_F}/K_{1_F} < 0.5$. Its anisotropy is about 4 times lower than that of the reference layer. For this reason, the in-plane magnetic field tilts the free layer magnetization away from the normal to the plane direction faster than the reference layer magnetization. Starting from zero field, for 0<H<2.5kOe, the in-plane magnetic field first yields a decrease in the relative angle between the magnetic moments in the two electrodes. Indeed, because the reference layer is initially oriented in a canted direction, $\theta_{c_R} \sim 10°$, the field-induced rotation of the free layer magnetization towards the applied field brings it closer to the reference layer magnetization. The minimum of resistance at H~2.5kOe therefore corresponds to the parallel orientation of both magnetic moments. Further increasing the magnetic field gives rise to an increase of the relative angle between the two moments so that correlatively the resistance starts increasing again. It is expected that at larger fields, the resistance would decrease again since the system would evolve towards the parallel configuration if full saturation could be reached at very large fields. But full saturation of the reference layer magnetization would require overcoming both the anisotropy energy and the antiferromagnetic RKKY coupling across the ruthenium layer. Field of the order of 2 T would be needed to observe this behavior which is out of our range of measurements.

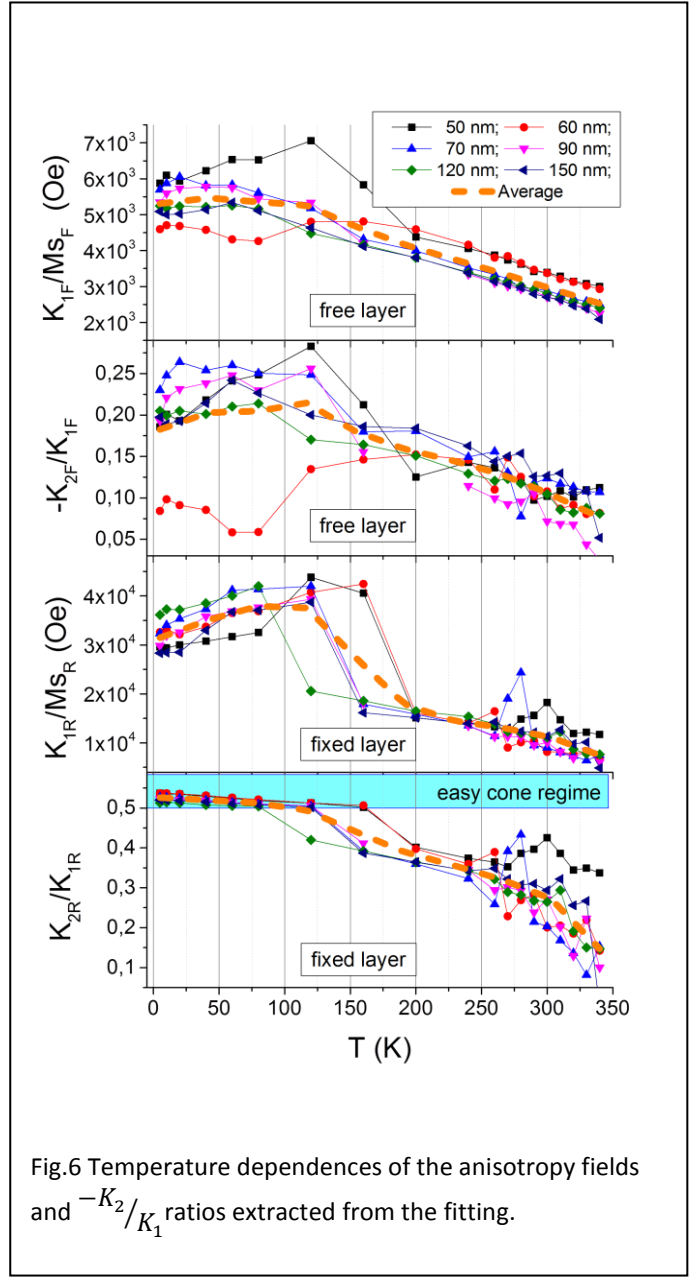

Fig.6 Temperature dependences of the anisotropy fields and $-K_2/K_1$ ratios extracted from the fitting.

A summarized view of the temperature dependences of $K_{1_F}/M_S$, $-K_{2_F}/K_{1_F}$, $K_{1_R}/M_S$, $-K_{2_R}/K_{1_R}$ extracted from the fitting for all measured devices is shown in Fig. 6. All samples exhibit the same trends and similar magnitudes of the anisotropy constants extracted from the fitting. Scattering of the extracted values gives an idea of the dispersion of the fitting parameters. The temperature dependences of average values of these parameters over all measured devices are also shown. For the free layer, in average, $K_{1_F}/M_S$ increases almost linearly as temperature lowers in the range 120K-300K. The corresponding $-K_{2_F}/K_{1_F}$ ratio also increases in this range of temperature, not exceeding 20% at low T and therefore never reaching the easy cone regime. Concerning the reference layer, the situation is generally similar, but $-K_{2_R}/K_{1_R}$ ratio is much larger at all temperatures. Below 120-160K, $-K_{2_R}/K_{1_R}$ ratio is above 0.5 so that the reference layer

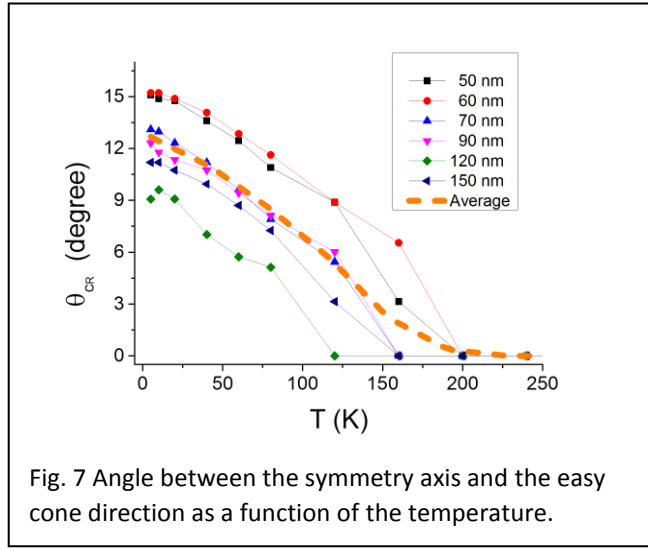

Fig. 7 Angle between the symmetry axis and the easy cone direction as a function of the temperature.

magnetization enters the "easy-cone" regime as pointed out above.

We also recalculated the easy-cone angle $\theta_{c_R}$ of the reference layer magnetization versus temperature. - As shown in Fig. 7, the easy-cone angle increases almost linearly as temperature decreases below ~180K. Furthermore, $\theta_{c_R}$ is observed to increase with decreasing diameter of sample. As will be shown further in section 7, the $K_2$ contribution is interpreted in terms of spatial fluctuations of the uniaxial $K_1$ first order term. In this case, for smaller diameter, edge defects may increase $K_2$ due to increased spatial fluctuations of $K_1$. This could explain the larger $-K_2/K_1$ ratio and correlatively the large easy cone angle observed at small pillar diameters.

6. Easy-cone regime in sheet FeCoB/MgO films

Generally, one cannot rule out *a priori* that certain type of micromagnetic distortions in the ferromagnetic electrodes could be responsible for the observed hard-axis MR(H) curve deformations in the studied pillars at low temperatures. To exclude this possibility, experiments were conducted at sheet film level in order to check whether the second order anisotropy is also evidenced in this case. In thin films, the demagnetizing (magnetostatic) energy and first order perpendicular interfacial anisotropy $K_{s1}$ have the same symmetry. They can be combined in one effective anisotropy density constant $K_1 = (\frac{K_{S1}}{t} - 2\pi M_S^2)$. Consequently, an easy way to tune the $K_2/K_1$ ratio simply consists in changing the thickness $t$ of the FM film. For any $K_2 \neq 0$ amplitude, a range of FM thickness around the anisotropy reorientation transition from out-of-plane to in-plane direction should always exist, wherein the easy-cone regime should be observable.

To check this, several samples were grown, consisting of an $Fe_{72}Co_8B_{20}$ layer in contact with MgO with nominal thickness of FM material $t$= 17.4 Å, 16.9 Å and 15.8 Å. Room-temperature magnetization measurements with magnetic field applied parallel to the films plane clearly exhibit three different M(H) loop shapes as shown in Fig. 8. The thickest and thinnest samples demonstrate M(H) loops respectively typical of XY-easy-plane and Z-easy-axis anisotropies (the field is applied in the XY-plane). The sample with intermediate FM layer thickness shows features of a two-step magnetization process. Firstly, an abrupt switching of magnetization, as in the thickest sample, followed by a slower non-linear M(H) magnetization increase. Such features are exactly expected in presence of easy-cone anisotropy. As discussed in the previous section, when the magnetic field is applied perpendicularly to the easy cone symmetry axis, it is initially very easy to rotate the in-plane component of magnetization around the easy-cone. This corresponds to the low-field part of the M(H) curve with an abrupt variation of the magnetization. Following this rapid rotation, at larger fields, the magnetization has to depart from the easy cone to gradually align with the in-plane applied field. This yields a more gradual increase of magnetization since the easy cone anisotropy has to be gradually overcome by the Zeeman energy. Corresponding macrospin simulations using model of Eq. (3) are shown in the inset of Fig.8 and reproduce the qualitative modifications of the in-plane M(H) loop shapes with the film thickness variation. Knowing that the "real" films are in multidomain state, we did not try to match the macrospin simulations and experiments exactly.

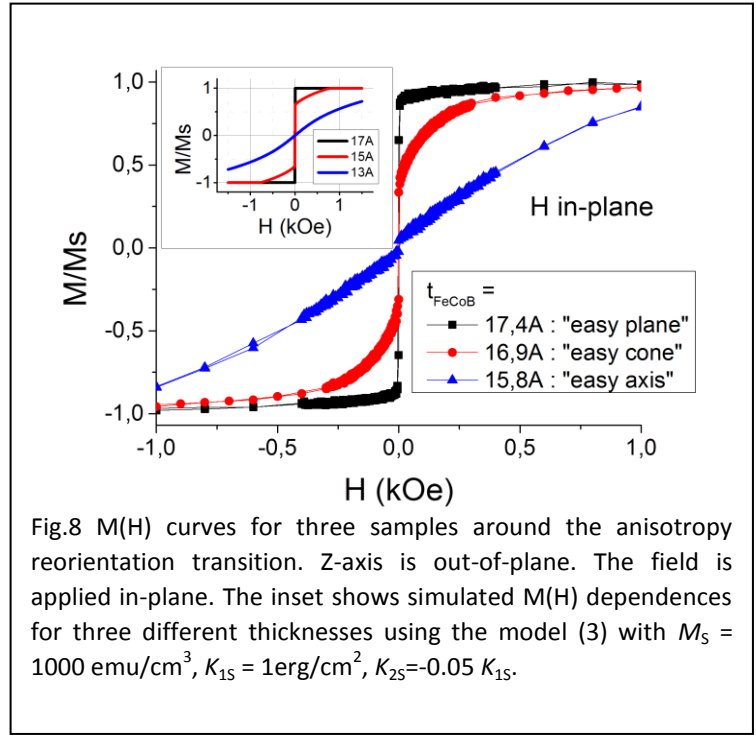

Fig.8 M(H) curves for three samples around the anisotropy reorientation transition. Z-axis is out-of-plane. The field is applied in-plane. The inset shows simulated M(H) dependences for three different thicknesses using the model (3) with $M_S$ = 1000 emu/cm$^3$, $K_{1S}$ = 1erg/cm$^2$, $K_{2S}$=-0.05 $K_{1S}$.

7. **Discussion**

Regarding the origin of the second order anisotropy term which gives rise to the easy cone regime, at least two explanations *a priori* can be provided and discussed. The first one is based on the possible existence of a bulk magnetocrystalline cubic anisotropy in the centered cubic Fe rich alloy constituting the magnetic electrodes of the MTJ. Our samples are polycrystalline so that the in-plane mosaicity of the FeCoB grains can average out the in-plane anisotropy. In contrast, due to the (100) texture of the film, the out-of-plane component of this anisotropy can be conserved with easy axis of anisotropy along the <001>, <010>, and <100> directions [24]. The four-fold bulk cubic anisotropy combined with the two-fold uniaxial anisotropies for the out-of-plane direction can yield the observed behavior for the M(H) dependence [25].

We have employed X-band magnetic resonance technique (9.45GHz) in order to investigate this possible source of 2$^{nd}$ order anisotropy in the sample with $t$ = 16.9 Å. Room temperature ferromagnetic resonance (FMR) spectra were measured for different angles of magnetic field with respect to the sample normal. The results are shown in Fig.9. FMR signal, as seen by comparing Fig.8 and Fig.9 is observed in a magnetic field range wherein the sample is completely saturated. The angular dependence is composed of four-fold and two-fold angular contributions of comparable amplitudes. These two contributions exhibit energy maxima when the field is oriented in-plane. Conversely, the four-fold anisotropy also reaches maxima when the field is out-of-plane while the two-fold anisotropy reaches minima for this field orientation. Not taking into the details of magnetic resonance, one can therefore definitely state that the hard axis directions of the four-fold anisotropy correspond to the normal to the film or to the in-plane direction. For the out-of-plane angular dependence of FMR, the expected behavior is qualitatively similar for the case of uniaxial + cubic anisotropy and for the case of uniaxial with the second order uniaxial term (see Appendix). The extracted constants in the case of uniaxial + cubic anisotropies are $K_1$=6.2·10$^6$ erg/cm$^3$ and $K_{1_C}$=-7.7·10$^4$ erg/cm$^3$ (assuming H rotating in (010) plane). In comparison with the bulk values of bcc iron (~5·10$^5$ erg/cm$^3$), the obtained cubic anisotropy constant $K_{1_C}$ is six times lower and has the opposite sign. It is well known that by adding cobalt into iron, the anisotropy constant $K_{1_C}$ is expected to gradually change from positive to negative with $K_{1_C}$~0 for Fe$_{45}$Co$_{55}$ composition [24]. In our case, the layer is iron-rich so that the lower value of $K_{1_C}$ could be explained by the Co content of the alloy in this 1.7 nm thick layer. However, the opposite sign of $K_{1_C}$ is not expected. Moreover, for positive $K_{1_C}$, the easy direction of uniaxial and cubic anisotropies would coincide along <001> direction not allowing therefore the formation of the canted state in contradiction with its experimental observation. We can therefore conclude that bulk cubic anisotropy of iron rich alloy does not play a significant role in these samples and other explanations have to be found for the second order anisotropy term.

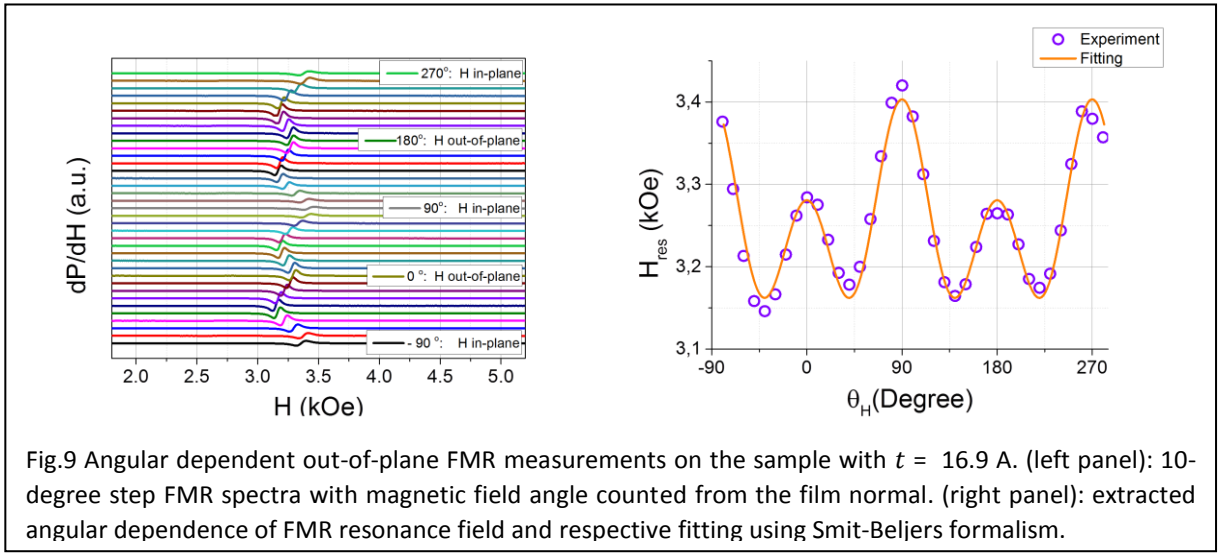

Fig.9 Angular dependent out-of-plane FMR measurements on the sample with $t$ = 16.9 A. (left panel): 10-degree step FMR spectra with magnetic field angle counted from the film normal. (right panel): extracted angular dependence of FMR resonance field and respective fitting using Smit-Beljers formalism.

Several experimental studies reported anisotropy reorientation phenomena about which the role of a second-order uniaxial anisotropy term could be evidenced. Easy-cone regime was observed experimentally near the magnetic reorientation transition in Co films grown on Pt(111) and Pd(111) substrates [16] as well as on Co/Pt multilayers [14]. Recently, J. Shaw *et al* have reported FMR measurements on Ta/Co$_{60}$Fe$_{20}$B$_{20}$/MgO films [19]. The authors obtained an angular dependence of FMR with maxima of FMR field corresponding to in-plane and out-of-plane magnetic field orientation with four-fold and two-fold angular dependencies, as in our case. The authors, however, used a cobalt-rich crystallized alloy, which in bulk has negative constants of cubic anisotropy [24]. They put forward a possible $\sim \sin^4 \theta$ contribution without too much explanation on its possible origin.

Besides, several studies have pointed out the possible influence of strain on second order anisotropy in thin magnetic films [26,27]. For instance in Ref [27], the authors observed an anomaly in the tunneling conductance in Ta/CoFeB/MgO based MTJ at low temperatures (T=160K), that they interpreted as a structural-magnetic phase transition of a magnetic oxide formed at the interface between MgO and CoFeB. Magnetic phase transition alters both spintronic and magnetic properties of the MTJ stack. While the proposed interpretation definitely needs a more detailed study, we should accept the fact that in our experiment we also observe a fast increase of $K_1$ for the reference layer in the temperature range 120-160K. It can be speculated that this observation might be associated with a low-temperature structural transition in one of the stack layers, not necessarily a magnetic one. Along the same line, mismatch of thermal expansion coefficients of the different materials in the stack and substrate can also play an important role. Considering the large magnetostriction of Fe rich FeCo alloys, stresses in the pillar can change the magnetic anisotropy in the magnetic layers through magnetoelastic coupling. Even the crystallization of MgO during the post-deposition annealing can produce some residual stresses in the neighboring ferromagnetic electrodes. Therefore one cannot rule out that magnetoelastic effects play a role in the second order anisotropy term that we observe in our samples. Further structural characterization and stress analysis would be required to clarify that.

Another possible origin of the second-order uniaxial term was proposed theoretically by B. Dieny and A. Vedyayev [20]. They have shown analytically that spatial fluctuations in the magnitude of first order surface anisotropy can give rise to a second order anisotropy contribution provided the characteristic wavelength of these fluctuations is much smaller than the exchange length. The sought second-order contribution has a negative sign with respect to the main first-order term, thus allowing the onset of easy cone anisotropy. Topology of the interface in their model determines the relative strength of the second-order contribution. In case of Fe/MgO systems, spatial fluctuations of the effective perpendicular anisotropy can be responsible for the second order anisotropy term. These fluctuations can be due to local variations in the ferromagnetic layer thickness associated with film roughness. Due to the competition between interfacial anisotropy and bulk demagnetizing energy, around the anisotropy reorientation transition, a monolayer variation in the thickness of the FeCoB layer due to interfacial roughness is sufficient to yield spatial variations of effective anisotropy from in-plane to out-of-plane. Following the model of Ref. [20], using an average film thickness of 15 Å and variations in FM layer thickness +/- 2 Å, one can expect spatial modulation of the surface anisotropy parameter of the order ~0.2 erg/cm². Considering an exchange constant ~ 1.5·10$^{-6}$ erg/cm, $K_{1S}$~ 1 erg/cm², period of spatial fluctuations ~ 15 nm, one should expect $K_{2S}$~-0.0024 erg/cm² which magnitude is much lower than that estimated from the aforementioned fits (Fig.5 and 6). Alternatively, one may think about the possible presence of nanometric "dead" spots where contribution to the net interfacial anisotropy could be locally strongly reduced. In Ta/CoFeB/MgO, a possible explanation for the existence of dead spots could be the preferential diffusion of Ta along the grain boundaries of the CoFe(B) layer to the MgO barrier upon post-deposition annealing. The presence of Ta next to the barrier can locally alter the interfacial perpendicular anisotropy yielding strong local variations of interfacial anisotropy between the inner part of the grains and the grain boundaries. Assuming a grain size ~ 16 nm and a spatial modulation of the interfacial anisotropy ~ 1 erg/cm² yields $K_{2S}$~0.07 $K_{1S}$ which is the right order of magnitude.

The observed temperature dependence can be explained by the model of Ref [20]. According to it, $K_2$ scales as square of $K_1$. This is generally what is observed in the temperature range 160K-340K: $K_1$ decreases with temperature but $-K_2/K_1$ also decreases, meaning that $K_2$ drops with temperature faster than $K_1$. However, in the temperature range between 5 and 120K, the behaviors of the reference and free layer are different. Indeed, the free layer keeps following the above described tendency while the reference layer shows abrupt changes in $K_1$ and further decrease of its magnitude versus decreasing temperature. This may indicate that for the reference layer which has a complex structure (SyAF), the macrospin description may not be sufficient. Different temperature dependences of perpendicular anisotropies arising from MgO/FM interface and from the synthetic Co/Pt multilayer as well as temperature-dependent coupling field through the NM layer may complicate the overall picture.

From practical point of view, the easy cone anisotropy can be used to significantly improve the writing performance of pMTJ-based STT-MRAM elements [28, 29]. In a standard pMTJ system, the magnetic moments of both free and reference layers are aligned parallel or antiparallel in standby regime. Upon writing, when the write current starts flowing through the MTJ, the initial STT-torque is zero and only thermal fluctuations or micromagnetic distortions provide the non-collinearity required to trigger the reversal of the storage layer magnetization. Both effects are generally undesirable in STT-MRAM technology. Indeed, thermal fluctuations are stochastic by nature and therefore the write pulse duration and intensity must be overdesigned to reach the specified write error rate. As for micromagnetic distorsions, the latter induce non-uniform switching process which can result in the need for higher switching current and variability in the switching process. An easy cone regime in the free layer and the easy axis configuration in the reference one would be the optimal configuration for a STT-MRAM memory element. Unique features of easy cone regime is that it allows for a canted state and at the same time conserves the axial symmetry so essential for effective transfer of the STT torque into the angular motion.

We point out that because the magnetostatic term reduces the effective $K_1$ but keeps $K_2$ unchanged, the ratio $K_2/K_1$ is at least twice larger than the surface constants ratio $K_{2s}/K_{1s}$. Thus, keeping $K_2/K_1 \sim 10\%$ as it is in the free layer, the expected $K_{2s}/K_{1s}$ ratio should not exceed 5% at room temperature which is quite easy to overlook both in experiments and theories.

**Conclusion**

While easy-axis magnetoresistance loops allow for determination of switching current and coupling fields, hard axis magnetoresistance loops provide additional information about the magnetic anisotropy in pMTJ pillars. Reversible parts of the hard-axis magnetoresistance loops starting from parallel or antiparallel configuration can be simultaneously fitted providing quantitative estimation of the effective anisotropy fields both in the free and reference layers.

In this work magnetoresistance loops of pMTJ pillars with radius 50-150 nm were measured in a wide range of temperatures. The anisotropy fields in both free and reference layers were derived in the temperature range between 340 and 5K. At temperatures below 160-120K, the shape of the hard-axis magnetoresistance loops changes qualitatively from parabolic to triangular which cannot be described by a model taking into account only first order magnetic anisotropy. By adding a higher order anisotropy term, the magnetoresistance loops could be fitted to the model over the whole temperature range and for all measured devices. The extracted anisotropy constants have shown that the second order term is noticeable and it has a negative sign with respect to the first order anisotropy term in both layers. At room temperature, the magnitude of the second order term is about 10% of the first order one in the free layer and about 20% in the reference layer. With decreasing temperature, the second order term contribution increases faster than the first order one and exceeds 50% of the first order term in the reference layer below 120K. This results in a change of the reference layer net anisotropy from easy-axis along the normal to the plane to easy cone. In this state hard-axis magnetoresistance loop acquires a triangular shape for the antiparallel branch and a double well shape with a maximum at H=0 for parallel branch. The free layer remains with a net easy axis anisotropy at all temperatures. Extracted temperature dependences of the anisotropy in both layers are quantitatively and qualitatively similar for all measured devices whatever their diameter. Therefore, the anisotropy transition from easy-axis to easy cone regime seems to be diameter-independent.

We have evidenced the existence of the higher-order term in simple FeCoB/MgO sheet films and it is experimentally accessible for the thicknesses corresponding to the magnetization reorientation transition. The Dieny-Vedyayev model proposed in Ref. [20] explains the second order magnetic uniaxial anisotropy contribution, $-K_{2s}\cos^4\theta$ with $K_{2s} < 0$, as a result of spatial fluctuations of the first order anisotropy parameter, $-K_{1s}\cos^2\theta$ with $K_{1s} > 0$. The preferred diffusion of Ta through the CoFe(B) layers towards the MgO interface upon post-deposition annealing and CoFeB crystallisation was proposed as a possible mechanism at the origin of these spatial fluctuations of the CoFeB/MgO interfacial anisotropy.

The canted (easy cone) state of the free layer could be advantageously used to improve STT writing performance in pMTJ pillars in STT-MRAM applications. Thus further research aiming at engineering high -K₂/K₁ ratio while keeping $K_1$ large enough to achieve sufficient thermal stability of the storage layer is highly desirable.

**Acknowledgements**

The authors acknowledge Dr. Sergei Nikolaev for useful discussions, Sergio Gambarelli for helping with FMR measurements. The work was partially supported by Samsung Global MRAM Innovation program and CEA-EUROTALENTS scholarship.

**Appendix**

By simple trigonometrical transformations, one can decompose both uniaxial including first and second order terms and uniaxial plus cubic anisotropy into the same $K_{2\theta}\cos(2\theta) + K_{4\theta}\cos(4\theta)$ dependence with effective constants $K_{2\theta}, K_{4\theta}$ which expressions will be different. In the case of uniaxial anisotropy including first and second order anisotropy terms, $E = -K_1\cos^2\theta - K_2\cos^4\theta$ is equivalent to $E = K_{2\theta}\cos(2\theta) + K_{4\theta}\cos(4\theta)$ provided $K_{2\theta} = -(K_1 + K_2)/2$ and $K_{2\theta} = -K_2/8$. In the case uniaxial plus cubic anisotropy $E = K_1\sin^2\theta + K_{1c}(\alpha^2\beta^2 + \beta^2\gamma^2 + \gamma^2\alpha^2)$ can be written as $K_{2\theta}\cos(2\theta) + K_{4\theta}\cos(4\theta)$ provided $K_{2\theta} = -K_1/2$ and $K_{2\theta} = -K_{1c}/8$ if the out-of-plane rotation of the magnetization takes place in the (010) plane of cubic anisotropy and provided $K_{2\theta} = -(K_1 + K_{1c}/4)/2$ and $K_{2\theta} = -3K_{1c}/32$ if the rotation takes place in the (110) plane. In case of annealed Fe/MgO interface, as it was already mentioned in the text, in-plane cubic anisotropies are averaged out due to the randomness in the in-plane orientation of the FeCo crystallites.